# Adaptive Artificial Immune Networks for Mitigating DoS flooding Attacks

Jorge Maestre Vidal[a],*, Ana Lucila Sandoval Orozco[a], Luis Javier García Villalba[a]

[a]*Group of Analysis, Security and Systems (GASS), Department of Software Engineering and Artificial Intelligence (DISIA), School of Computer Science, Office 431, Universidad Complutense de Madrid (UCM), Calle Profesor José García Santesmases s/n, Ciudad Universitaria, 28040 Madrid, Spain*

## Abstract

Denial of service attacks pose a threat in constant growth. This is mainly due to their tendency to gain in sophistication, ease of implementation, obfuscation and the recent improvements in occultation of fingerprints. On the other hand, progress towards self-organizing networks, and the different techniques involved in their development, such as software-defined networking, network-function virtualization, artificial intelligence or cloud computing, facilitates the design of new defensive strategies, more complete, consistent and able to adapt the defensive deployment to the current status of the network. In order to contribute to their development, in this paper, the use of artifi- cial immune systems to mitigate denial of service attacks is proposed. The approach is based on building networks of distributed sensors suited to the requirements of the monitored environment. These components are capable of identifying threats and reacting according to the behavior of the biolog- ical defense mechanisms in human beings. It is accomplished by emulating the different immune reactions, the establishment of quarantine areas and the construction of immune memory. For their assessment, experiments with public domain datasets (KDD'99, CAIDA'07 and CAIDA'08) and simulations on various network configurations based on traffic samples gathered by the University Complutense of Madrid and flooding attacks generated by the

---

*Tel. +34 91 394 76 38, Fax: +34 91 394 75 47

*Email addresses:* jmaestre@ucm.es (Jorge Maestre Vidal), asandoval@fdi.ucm.es (Ana Lucila Sandoval Orozco), javiergv@fdi.ucm.es (Luis Javier García Villalba)

tool DDoSIM were performed.

---

## 1. Introduction

By definition, Denial of Service (DoS) has the objective of disabling computer systems or networks. The DoS attacks with origin in multiple sources are referred as Distributed Denial of Service (DDoS) attacks. In recent years, the number of incidents related with these threats reported by the various organizations for cyber defense shows an alarming growth. According to the European Network and Information Security Agency (ENISA), between 2013 and 2014 an increase of 70% was observed (Marinos, L. and Sfakianakis, A. (2016)). In addition, they pose a threat that has begun to be used in order to achieve other objectives. These include disguising activities in relationship with malware spreading, concealment of fraudulent money transfers (DoS (2016d)) or compromising anonymous networks, such as Tor or Freenet (Jansen et al. (2014)). This growth is attributed to various reasons: the first of them is that DDoS attacks are usually triggered by previously infected systems, which in most of the cases are part of botnets. The botnets have been adapted to be resilient against the classical detection schemes, thus allowing the construction and maintenance of larger collections of zombies and increasing their difficulty to be identified. (DoS (2016f)). Another im- portant reason is that attackers are able to take advantage of amplifying elements, in this way enhancing their potential to be harmful. To do this, they exploit vulnerabilities in protocol implementations on the intermediate network devices, particularly at DNS, NTP and SNMP. On the other hand, as the European Police Office (Europol) warns (DoS (2016g)), the DDoS is becoming increasingly linked with the organized crime. Rent botnets for execution of these attacks is a very profitable business on the black market, often supplied as Crimeware-as-a-Service (CaaS). Finally, offenders with lack of formation have a wide variety of tools for easily configuration and deployment of flooding attacks. The black market also offers technical support, a situation that expands the range of user profiles which are able to attack with success.

The most common DDoS methods are based on flooding. Because of this, they are the principal object of study in this research. The flooding attacks modus operandi involves the injection of large volumes of traffic in order to saturate the victim systems (Zargar et al. (2013)). The popularity of this group of DoS attacks is mostly due to cheap price and simple implementation, in comparison with the good results they provide. Thus, there are lots of proposals aimed at their detection and mitigation. However few of them meet all the requirements to be effective in real use cases, emphasizing the needs of high true positive rates, unrepresentative false positive rates, low consumption of computational resources and real-time performance. In addition it is noteworthy that proposals of the literature seldom consider advantages of the new trends on networking. It is expected that the emerging networks, taking the example of 5G, increasingly move towards self-management. The use of novel technologies such as Software-defined networking (SDN), Network-Function Virtualization (NFV), Artificial Intelligence or Cloud computing, facilitates the design of Self-Organizing Networks (SON). This should encourage the appearance of more interesting proposals that are capable of reacting with a much more comprehensive view of the problem being treated.

To serve this cause, a strategy for detection and mitigation of DDoS flooding attacks is proposed. Therein the deployment of a sensor network that integrates an Artificial Immune System (AIS) inspired by the biological defense mechanisms of human beings is introduced. Unlike similar proposals, conventional bio-inspired methods for pattern recognition were not applied. Instead a combination of strategies for DDoS detection based on the study of variations of the entropy on the network traffic by thresholding, with the adaptation of the biological immune reactions is proposed. This makes it possible to apply real-time countermeasures, building an immune memory and establishment of quarantine areas, all in accordance with the current state of the protected network. In view of this, it is important to highlight the two major contributions of this paper: firstly, a new method for detecting DDoS that is able to forecast anomalies on the entropy of the traffic analyzed, and thereby recognition of flooding attacks is introduced. This is performed by representation of the entropy in time series and the definition of prediction intervals. On the other hand, a strategy for management of immune agents that implement the previously described detection system is proposed. Within this, the decisions are made as to when and how they will act and in what level of restriction, all this depending on the status of the network and orchestrated by an artificial immune approach.

The proposal was implemented and evaluated in different experimental scenarios, which involved public domain datasets (KDD'99 (DoS (2016e)),

CAIDA'07 (DoS (2016b)) and CAIDA'08 (DoS (2016a))), simulations on various network configurations, traffic monitored in the University Complutense of Madrid (UCM) and flooding attacks generated by the tool DDoSIM (DoS (2016c)).

The paper is divided into seven sections, and the first of them is the present introduction. The background necessary for a better understanding of the approach is described in section II. The proposed AIS is introduced in section III. The novel DDoS detection method implemented in the vari- ous agents of the AIS is detailed in section IV. Experiments, datasets and methodology are described in section V. Results are discussed in section VI. Finally, conclusions and future work are presented in section VII.

## 2. Background

The following describes all aspects necessary for understanding the proposal. Among them it is important to highlight those involving the characteristics of the DDoS flooding attacks and their countermeasures, a general review of the human being immune system and the different approaches with this in mind, in order to provide defenses against cybercrime.

### 2.1. Flooding threats: attacks and countermeasures

According to (Wei et al. (2013)), there are two types of traffic injection able to compromise a system or network by flooding. The first one is based on the constant and continuous generation of large volumes of information, and is well known as high rate flooding. This is a method which is usually very visible that easily overflows the computing capacity of the victim. On the other hand, the victim may be compromised by less noisy attacks, which are able to exploit vulnerabilities in the various communication protocols. They are known as low rate flooding attacks, using as a typical example, the attacks with On/Off patterns addressed against the TCP (Tang et al. (2014)). In both cases, the malicious traffic may be sent to the victim in a direct or reflected way (Bhuyan et al. (2015); Anagnostopoulos et al. (2013)). Most efforts of the community to deal with denial of service assume these behaviors, and from them different methods for detection, mitigation and identification of sources are proposed. Their most important features and evaluation schemes are described below.

The detection of the attacks is often necessary to conduct either of the other defensive reactions. This is based on the analysis of traffic that flows through the protected environment, looking for signatures of previously known attacks or anomalous behaviors. For that purpose various techniques have been proposed, such as probabilistic models based on Markov (Shin et al. (2013)), Genetic Algorithms (GA) (Lee et al. (2012), Chaos theory (Chen et al. (2013)), CUSUM statistical analysis with wavelet transforms (Callegari et al. (2012)), forensic methods based on visualization (Cai et al. (2010)), fuzzy logic (Kumar and Selvakumar (2013)) or the study of variations on the traffic entropy (Ozcelik and Brooks (2015); Bhuyan et al. (2015)). In approaches like (Zhou et al. (2014)), the problem of the similarity of the nature of the DDoS attacks in comparison with non-malicious events is studied. This is the case of the well-known situations such as flash crowds, which often occur when a large amount of legitimate users converge on a certain service in a short time interval.

The total or partial reduction of the damage inflicted by the attacks is defined as mitigation. To do this, it is common to use honeypots (Heckman et al. (2013)), puzzles that recognize non-human users (Zhu et al. (2014)), bandwidth enlargement (Khanna et al. (2012)), filtering or the adoption of security protocols such as IPsec (DoS (2016g)). As can be observed, the set of mitigation actions may also contain prevention strategies. These are characterized by not having direct dependence on the attack detection.

To find the compromised systems from which the malicious traffic is originated is referred to as the identification of their sources. Ideally, its objective is to track the cybercriminal. However, given the administrative difficulties that this process involves (different Internet Service Providers (ISPs), proxies, data privacy legislations, etc.), and the recent advances on footprint occultation, this goal is often very hard to carry out successfully. Consequently, many of the proposals in the literature just focus on getting as close as possible to the attacker, in order to sanitize the largest amount of regions within the protected network. On the identification of sources, the packet traceback is the most frequent approach. In (Alenezi and Reed (2014)) this issue is discussed, a lot of current approaches are collected, and a new scheme for uniform tracking is proposed. The Passive IP Traceback (PIT) that bypasses the deployment difficulties of the conventional IP traceback techniques by investigation of ICMP error messages is proposed in (Yao et al. (2015)). Finally, in (Jeong and Lee (2014); Yao et al. (2015)) the influence of the characteristics of the network topology on the effectiveness of the strategies for packet marking is studied.

The evaluation of defense systems against DoS/DDoS is a controversial issue today. Over the years, various collections of traffic samples and methodologies for assessment of these tools were proposed. They are mainly based on public domain datasets containing both legitimate traffic and DoS/DDoS threats. The malicious content usually corresponds with real traffic captures (KDD'99, DARPA'99, FIFA World Cup'98, etc.) or traffic generated by tools that imitate the behavior of the real attacks (D-ITG, Harpoon, Curl-loader, DDOSIM, etc.). In (Bhatia et al. (2014)) each of them is discussed in depth, and the lacks on the current verification methods are summarized. Thus they conclude that the traffic captures which were traditionally applied are outdated and widely disparate of the current traffic. For this reason, it is rec- ommended to consider only CAIDA 07 (DoS (2016b)) among them, labeled as the best of the bad solutions. On the other hand, the use of simulation tools entails the loss of realism. Also it is difficult to compare new proposals with those on the bibliography (Kumar and Selvakumar (2013)). This is be- cause every paper defines their own scenarios of experimentation according to their requirements.

## 2.2. Human beings immune system

All living beings have developed multiple immune mechanisms, emphasizing among them defenses of vertebrate species due to their sophistication. Many types of proteins, cells, organs and tissues form part of these systems, and they are related through an elaborate and dynamic network. As part of this more complex immune response, the human immune system, over time, adapts to recognize specific antigens, which is called adaptive immunity. The defense mechanisms compose the innate immunity, and usually are the first line of protection.

Each component of the innate immunity is able to recognize and eliminate different types of antigens. They are mainly rejection reactions conducted by external physical barriers such as skin or mucous membranes, and internal defensive elements like Natural Killer (NK) cells or phagocytes. They lack immune memory, and are only effective against pathogens known as priori.

On the other hand, the adaptive immunity presents specificity, i.e., after learning how to identify and reject an antigen, the knowledge gained allows it to react more firmly against the intruder, by generating new and stronger agents; but they can only act for mitigating the threat for which they were created. The most important cells involved in this process are lymphocytes and presenting cells, and the most important adaptive immune responses are humoral and cellular. In both of them take part agents responsible for antigen recognition and disposal. Given they pose a good example of this process, are briefly explained below.

- *Humoral response*. The antibodies detect and eliminate the threats by swallowing them. The remains of this process are captured by $T_h$ lymphocytes, and these stimulate the $T_b$ lymphocytes to generate an even greater amount of antibodies specialized in recognizing the threat. Antigens never seen before are identified by the antibodies that have suffered small mutations in their construction process, allowing them to recognize different antigenic determinants.

- *Cellular response*. The $T_h$ detect the intrusions. Then they attract $T_c$ cells for disposal. As in the previous case, the remains stimulate the generation of $T_h$ and $T_c$, specializing them against the new threat. In addition, the recent $T_c$ are often able to identify the antigen.

In both reactions, increasing defensive measures is temporary. After a quarantine period, the system is regulated by removing the excess of agents. These are programmed to die by apoptosis, also known as cell death.

Acquired immunity is the basis of the vaccination in human beings. When samples of an antigen are detected, the organisms deploy temporary and specific countermeasures for disposal. Therefore the response is faster and more effective than in the first contact.

### 2.3. Artificial Immune Systems

The adaptation of biological defenses towards information security is usually performed by deployment of multi-agent systems (Ou (2012)). In pioneering approaches, such as (King et al. (2001); Harmer et al. (2002)), the main guidelines for the emulation of the activities carried out by immune cells were introduced. They often apply some of the four classical bio-inspired algorithms: negative selection, clonal selection, artificial immune networks and Danger Theory (DT), which are briefly described below.

Negative selection is the process by which the immune agents learn to distinguish antigens from the cells of the organism itself. In (Zhaowen et al. (2012)) there is a good example of its application for detection of anomalies. But as the authors suggest, it poses a methodology that tends to generate high false positive rates, a situation that often leads to its complementation by clonal selection algorithms (Ligeiro (2014)). These are based on the assumption that every lymphocyte at its growth stage must be able to react against a specific antigen before being released in the body. Clonal selection is an effective solution to problems of recognition and optimization (Castro and Zuben (2002)). In proposals like (Sheshtawi et al. (2010)), it is also applied for identifying malware, often assuming an important refining step over negative selection.

The immune networks stem to the idea of extending clonal selection to networks. They are commonly implemented as a channel of interaction between the different actors of AIS. In (Seresht and Azmi (2014)) there is an example of an immune network for connecting different agents that perform negative selection. In (Yang et al. (2011)) a similar deployment is adopted, but this time involving agents that apply Danger Theory.

The Danger Theory takes into account the latest advances in medical research. Consequently and unlike their predecessors, it rejects the idea that organisms have the capacity to distinguish between own cells and antigens. Instead it postulated that the triggering of immune reactions is originated by warning signals sent from tissues in direct contact with the threat. Because of its novelty, it is one of the most common algorithms in the bibliography of recent years. In (Aickelin et al. (2003)) the bases for its adaptation to intrusion detection are defined. DT is applied to recognition of enumeration attacks in (Greensmith et al. (2010)), and it is also considered for intrusion detection based on studying the system calls of the protected environment in (Azmi and Pishgoo (2013)).

Despite the predominance of these methods in the bibliography, not all AIS are based on such specific processes of the biological immune systems. Some approaches imitate the global behavior of the innate and adaptive immune responses on the vertebrate beings, without following predefined algorithms. This makes it easier to combine the defensive strategies used in each field, with the ideas provided by biological immune systems, by this way reaching solutions that best fit the real use cases. A good example of this is proposed in (Boukerche et al. (2007)), where this idea is implemented to detect anomalies in network traffic. To do this, six main classes of agents are considered, among which are distributed sensing, communication and reaction tasks. Their adaptive immune response involves the cloning of a greater number of antibody agents in the threatened regions, by this way increasing their presence on the protected environment. In (Chen (2010)) antibodies are mobile agents that swarm the network looking for indicators of damage. Another example is (Visconti and Tahayori (2011)), wherein when the immune agents identify a potentially harmful incidence, the adaptive reaction is triggered. This entails the increase of the weights of the monitored features, thus gaining restrictiveness and specificity. Finally, in (Venkatesan et al. (2013)) a solution to some of the attacks that affect mobile networks is proposed. As in previous approaches, its behavior is based on cloning the agents that have been able to recognize a specific threat.

## 3. Biological immune reactions against DDoS

To design a bio-inspired system for detecting and mitigating denial of service attacks requires being aware of certain aspects of both threats to be treated and features of the defensive deployment. In order to establish the limitations and goals of this approach, the following describes the most relevant assumptions and requirements that have been taken into account throughout its development.

- The denial of service attacks to be treated may originate in one or more sources. In addition, they acquire the ability to take alternative routes to the victim, adapted to network conditions, such as filters that drop malicious traffic or congestion.

- Malicious traffic may harm the protected network in different places: proximities to victim nodes, intermediate network actors, or close to their source. Once a section is compromised, it is possible that a chain reaction leads to the compromise of the others.

- When a region on the protected environment is victim of a flooding attack, it must be quarantined until any hint of aggression disappears. In this way it is possible to provide proactive perimeter security.

- Although a region is in quarantine, the legitimate traffic should continue flowing through it practically in its normal rhythm. In other words, the quarantine should not block sections of the network nor prejudice their quality of service in a representative manner. Other- wise, the attacks achieve their goal.

- Mitigation actions should be performed as close as possible to the source of the threat. Thus the amount of network regions at risk is reduced, and the identification of the attacker is facilitated.

- Communications between the various immune agents must be performed through secure channels, in this way preventing that the attacker exploits them.

- Immune agents must be able to be activated/deactivated according to the state of the protected network.

- Defensive action must be proportionate to the risk to be treated. Thus the impact of the autoimmune reactions triggered by false positives is reduced.

- In order to enhance the alert correlation processes and the tasks performed by human operators, a record must exist which collects the current state of the network, the different immune reactions and their effectiveness.

As can be observed, these premises meet an important part of the needs of the current networks. However there are other aspects that have been set aside, highlighting among them the fight against the various evasion strate- gies. These contain methods for disguising the source of the attacks or hinder the tracking of the flooding path (Fast-Flux, Domain Generation Algorithms (DGA), exploitation of anonymous networks, etc.) (Zargar et al. (2013); Al-Duwairi and Al-Hammouri (2014). On the other hand, there are algorithms designed for misleading the detection system and thus do not triggering countermeasures, such as those proposed in (Ozcelik and Brooks (2015)). Their consideration involves adding a lot of complexity to the proposal, and because of this, it is out of the scope of the paper. Another important aspect that is also delegated to future work is to facilitate the interoperability with security protocols and data protection policies. We understand that analyz- ing obfuscated headers also means an important increment on its complexity, not being recommended for a first approach. Bearing this in mind, the architecture, behavior and properties of the proposal are described in depth below.

### 3.1. Architecture

The proposed system has distributed architecture and its different actors assume the various roles of the biological immune systems. Its success depends mainly on two types of agents spread along the protected network: $H$ detectors ($D_H$) and $A$ detectors ($D_A$). $D_H$ are involved in the innate immune response and in the adaptive response. Consequently they are capable of recognizing and blocking new attacks, and they participate in the construction of the immunological memory. On the other hand, $D_A$ have the ability to detect and mitigate the attacks previously identified by $D_H$ assuming a very important role in the adaptive response.

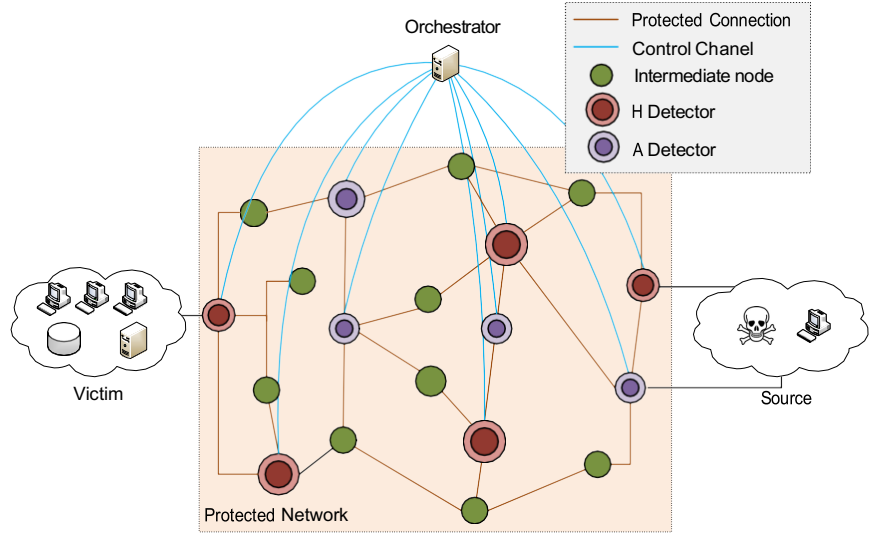

Figure 1: Example of distribution of the different actors in the AIS

In Fig. 1 an example of their deployment is shown. Apart from $D_H$ and $D_A$, other important components take part in the immune process, which are defined below.

- *Protected network*. The protected network is defined as the network segment that connects the attacker with the victim, and where the AIS is operating. Therefore, it is the element to be protected and the main scenario of different immune reactions.

- *Intermediate nodes*. Every node on intermediate subnets between the attacker and the victim that is not capable of detecting, mitigating or taking part in the immune reactions is defined as intermediate node.

- *H detectors*. The immune system agents that perform innate reac- tions, and are capable of triggering adaptive reactions are defined as $H$ detectors or $D_H$.

- *A detectors*. The immune system agents that only perform adaptive reactions are defined as $A$ detectors or $D_A$.

- *Orchestrator*. The orchestrator is the mediator between the immune agents. Its most important tasks are delimitation of the scope of the activation signals triggered at the immune responses, management of quarantine areas, separation of the AIS control data with respect to the protected network traffic and information gathering.

- *Protected connection*. Every connection between nodes on the protected network over which the proposed AIS acts is defined as protected connection.

- *Control channel*. The connections between the orchestrator and immune agents are referred as control channel.

The distribution and cooperation of the various components of this architecture allows performing innate and adaptive responses to attacks. The deployment of an orchestrator on a different data plane from the protected network connections provides autonomy, prevents that the traffic generated by the AIS penalize QoS, and facilitates the adoption of security protocols, thus reducing the risk of packet poisoning or Man-in-the-Middle attacks. But despite its obvious benefits, it is optional, because in certain use cases it is not possible to have much control of the protected environment (law restric- tions, privacy, etc.). In this situation, the immune agents may have sufficient autonomy to do without it, which involves: being in direct contact with agents over which it has influence, saving a log with the performed actions and determining whether it belongs to a region in quarantine or not (and the consequences that this entails). Usually these actions can be carried out easily, as for example by the implementation of tunneling between agents for their communication, the use of counters to determine the quarantine period, etc. but not with the global overview of the protected network and the advantages that the orchestrator provides.

### 3.2. Artificial immune responses

Each response conducted by the proposal is subject to conditions and events that emulate the behavior of the biological immune systems. With this in mind, the behavior of the AIS is mainly conducted by two different artificial immune responses: innate and adaptive.

### 3.2.1. Innate response

As in biological systems, the innate immunity on the approach is the first line of the defense strategy. It aims to identify and mitigate new threats and protect $H$ detectors of disablement by flooding. The process of innate immunity requires maintaining activated $D_H$ agents along the protected network. These sensors monitor the entire traffic flowing through them looking for suspicious anomalies. Therefore, detected attacks must present certain evident characteristics related to considerable fluctuations in the analyzed traffic distribution. Once a threat is identified, the mitigation measures consist mainly on the adoption of directives that restrict the communications with nodes, ports or services involved in the attack vector. We are aware that more sophisticated mitigation actions could be implemented, but their decision and development are delegated to future studies in order to facilitate a better understanding of this first approach.

The innate response provides quick and efficient countermeasures, requiring no communication with the orchestrator prior to their launch. By recognition and elimination of pathogens before they enter into the system, the proposal innate response emulates the behavior of the immune system of human beings. This is because it acts in the same way as the various external physical barriers or cells, and without specificity. In addition, it should be noted how agents involved in this task act coincide with those of most conventional Intrusion Prevention Systems (IPS), i.e. IDS with the ability to apply basic countermeasures.

### 3.2.2. Adaptive response

The adaptive response is the next defensive step in the proposal. It is triggered every time a $D_H$ agent recognizes a new threat, which implies that they must hold at least a short memory capable of storing their latest decisions. In this context, determining when an attack is Non-Seen-Before (NSB) im- plies it is not presence in the immune memory. Because of this, that memory has a very important role in the AIS decision-making. Determination of how the immune memory will be implemented is not usually a trivial problem. With this purpose, the two more intuitive schemes to take into account are those centralized or distributed. When the immune memory is centralized, is sustained by the orchestrator. In this case this component is responsible for determining whether an incidence is NSB, considering information provided by all the sensors. However, if it is distributed, each detector disposes only the information gathered for itself, or by a group of close sensors. Hence an attack tagged as NSB by an agent could not be new for other detectors. This second approach provides greater autonomy to the agents and allows some of them to trigger different adaptive reactions against the same pathogen. Given that this is a more efficient strategy and a good way of covering separated regions of the protected network, it was implemented at the experimentation.

Once the adaptive reaction is released, the $D_H$ that identified the attack sends activation signals to the $D_A$ agents in close proximity. The scope of these signals can be determined in different ways. It may affect the surrounding neighbors, nodes on a certain region or network devices in specific routes previously established by the orchestrator. This also enables the application of Artificial Intelligence and Data Mining in order to research their optimal propagation, giving many possibilities for future works. For simplicity the first of the previously mentioned alternatives was implemented.

Then the activated $D_A$ agents analyze traffic flowing through them. Unlike $D_H$, their detection engines increase restrictiveness in proportion to the flood of the attack, usually acting much more stricter than $D_H$. In this way it is prevented that the division of the attack flow reach the victim by alternative routes, assuming that when it is split, it becomes less noisy and hence more difficult to be detected. In order to prevent this measure result- ing in a substantial increase in the false positive rate, specificity is taken into account. To ensure specificity, they are only able to apply countermeasures against the threat that has activated them. Therefore, they can only take action against several attacks if they have been activated to mitigate each of them. Because of all of this, and as in nature, the artificial adaptive immune response involves the increase of the amount of effectives able to react against a certain triggering attack.

At the end, the deployed countermeasures are effective for a certain period of time. While the threat persists, the immune response remains activated. If it is no longer visible, a quarantine period is activated. The quarantine is interrupted only upon detection of replicas of the intrusion (implying back to the previous state), or when the countdown expires. The network segments covered by a set of $D_A$ agents active against a specific threat and coordinated by the same $D_H$ sensor, are their quarantine region.

### 3.3. Implementation

The behavior of the artificial immune responses is implemented in the following procedure:

1. At first, $D_H$ analyze all the traffic flowing through them. $D_A$ remain on standby waiting to be activated.

2. When the $D_H$ identify malicious traffic, the related connections will be blocked. This is an innate immune response. Then, the sensor warns the $D_A$ surrounding to proceed with the activation of countermeasures. In this way the adaptive immune response is triggered. The warnings contain information about their origin and the malicious flow characteristics.

3. When activated, the $D_A$ analyze the traffic from their potentially harmful sources. Unlike in $D_H$, the level of restriction of their decision threshold are increased according to the characteristics of the triggering attacks. They also block the identified threats. Thus, if the attacks cross alternative paths, they are also mitigated.

4. In the case that different $D_H$ emit activation signals for the same attack, the $D_A$ apply the most restrictive one.

5. The $D_A$ return to standby mode just in case a period of time without news of the attacker has passed, or triggering signals have not been received.

6. $D_H$ remain vigilant to face new incoming threats.

An example of the behavior of the proposed AIS when dealing with DDoS flooding attacks is described in Fig. 2. It is part of the situation shown in Fig. 2a, where S is the source of the attack, $T$ is the target, and each $N_i$ is an intermediate node located at $i$. In the case that the sensor $D_H$ is unable to recognize the threat, it is propagated along the protected network via different routes and according to the load balancing policies, as shown in Fig. 2b. But if the attack is successfully detected, the innate immune response is initiated. Thus the traffic from $S$ is discarded, slowing the advance of the intrusion, as shown in Fig. 2c. As a legitimate flow trying to reach its destination when some network incident block its route, the malicious traffic will try to reach the victim for alternative paths. Fortunately, when the threat was detected the adaptive immune response was also triggered. As shown in Fig. 2d, this has led to the activation of the neighboring $D_A$ agents. Consequently, the attack reaching the victim through the new connections is prevented.

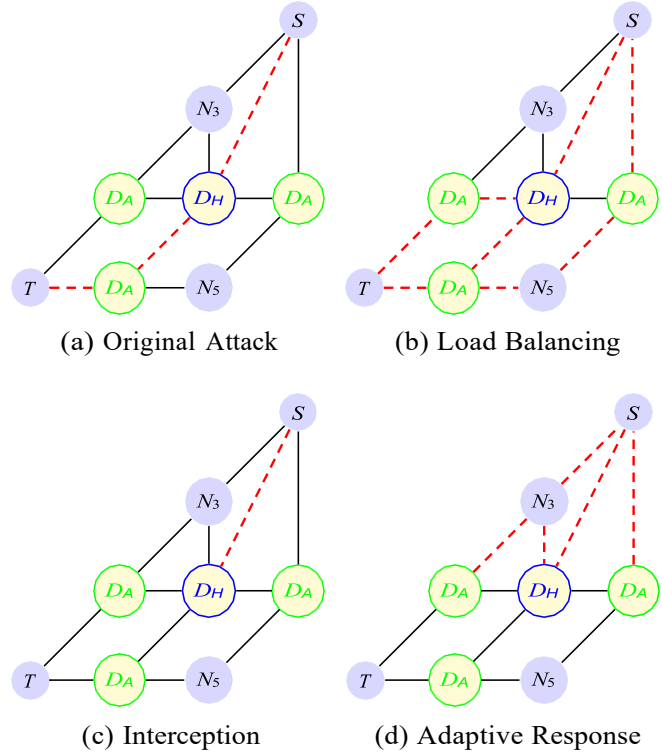

(a) Original Attack

(b) Load Balancing

(c) Interception

(d) Adaptive Response

Figure 2: Example of the AIS behavior

Despite its simplicity, this scheme has a large number of advantages. Firstly, it increases the defensive measures when a threat is recognized, proportional to the risk level. In addition, the specificity makes it only occur on specific connections, without affecting other traffic routes and reducing the impact of the false positives. Furthermore, the presence of two types of agents allows it to strategically adapt the defensive network to the monitor- ing environment. In this way the most powerful computers may act as $D_H$, and the rest assume the role of $D_A$. Note that a single node can assume both roles, or even deploy their own virtual network including several of them. On the other hand, the restriction level is self-regulated through deactivation of agents when the quarantine time passes, thus saving resources.

*3.4. Properties*

The main characteristics of the biological immune systems have been assumed and adapted to the fight against DDoS in the following ways:

- *Innate and adaptive responses*. At the innate response, the proposal has the capacity to react against threats which were not previously recognized. Once detected, it strengthens the defensive measures against future replies as occurs in biological adaptive immune reactions.

- *Specificity*. In nature each adaptive immune cell reacts against only one type of antigen (unlike the case with innate responses). Bearing this in mind, in this approach the innate response is common to any type of threat detected. However, the immune response affects only the triggering flooding attack.

- *Clonality*. When unknown antigens are detected, the adaptive biological response clones the cells that have been able to be recognised. This increase the chance of identifying their replicas. The AIS behaves in a similar way when activating agents after the adaptive response is released.

- *Immune memory*. Both in nature and the proposal, the deployed countermeasures remain active for a period of time. This enables them to react more effectively against future replicates.

- *Self-Regulation*. Once the adaptive response is triggered, the human immune system is regulated by the apoptosis of most of the new cells. In the AIS, defensive measures are also reduced when a certain period of time is exceeded and no replications of the triggering threat are detected.

- *Autonomy*. In both cases, once the entities of the defensive deployment are activated, they operate with complete autonomy.

- *Diversity*. In nature, the set of immune agents must be able to detect any antigen. The same thing happens in the AIS, where specificity is applied only in the deployment of countermeasures.

## 4. Detection of flooding attacks

The following describes the most important aspects of detection of threats and identification of sources on the approach, which involves metrics, forecasting, thresholding and recognition of compromised nodes.

### 4.1. Metric

The monitored traffic is analyzed by studying the entropy of its distribution. In particular, entropy fluctuations are analyzed looking for anomalous behaviors. The decision to use entropy variations over other detection methods proposed in the bibliography is that, as demonstrated in (Ozcelik and Brooks (2015)), is much more effective. This is principally because its accuracy depends less than those of the others on how the protected network is used. The traditional entropy was first adapted to the information theory by Shannon in 1948 (Shannon (1948)). It was considered as a measure of fluctuations on qualitative variables, and it is often defined as the degree of unpredictability on their behavior. Given the qualitative variable $X$, the finite set $\{x_1, x_2, \cdots, x_n\}$ and their probabilities $p_1, p_2, \cdots, p_n$, the Shannon entropy was described by the following expression:

$$H(X) = \sum_{i=1}^{n} p_i \, \log_a \frac{1}{p_i} = - \sum_{i=1}^{n} p_i \, \log_a p_i \qquad (1)$$

where $\log_a b \times \log_b x = \log_a x$. In addition, if the variable is deterministic then $H(x) = 0$ must be satisfied. This means that all the $p_i$ probabilities are 0, except for one, which has value 1. There are different generalizations of this entropy, adapted to the different use cases. The AIS applies that proposed by Rènyi. This decision was made considering studies like (Bhuyan et al. (2015)), where its effectiveness stands out from the rest of variations when applied on the detection of DDoS flooding attacks. Rènyi entropy is defined as follows:

$$H(X) = \frac{1}{1-\alpha} \log_2 \sum_{i=1}^{n} p_i^{\alpha} \qquad (2)$$

where the parameter $\alpha$ indicates its order, such as $\alpha \geq 0$ and $\alpha \neq 1$. Note that the Shannon entropy is the particularly case proposed by Rènyi when $\alpha = 1$.

In our approach, the variable $X$ is defined as the volume of traffic flowing between two network nodes $A$ and $B$, destined to port $C$. The $P$ probability implies that every $p_i$ represents the frequency of occurrence in the monitored traffic of packets with certain source, destination and that are addressed to a specific port. Therefore it is fulfilled that:

$$p_i = \frac{a_i}{No.packets} \tag{3}$$

where $a_i$ is the total amount of packets that met the previously described condition. In this proposal the entropy variations are treated as univariate time series of $N$ observations, expressed a:

$$H_\alpha(X)_{t=0}, \ H_\alpha(X)_{t=1}, \ \cdots, \ H_\alpha(X)_{t=N} \ ; \ (H_\alpha(X)^N{}_{t=0}) \tag{4}$$

### 4.2. Forecasting entropy variations

When estimating the entropy of future observations it is taken into account that $H_\alpha(X)^N{}_{t=0}$ series may experiment changes in trend and seasonality over time. Additionally, the forecasting methods to consider should be effective with few observations, and able to run efficiently in real time. For this reasons, to model the network behavior the triple exponential smooth- ing proposed by Holt-Winters has been chosen. This election is supported by publications like (Gardner and Dannenbring (1980); Groff (1973)), where it is shown that considering the trend and seasonality of series in which they are unrepresentative, leads to insignificant prediction errors. Furthermore, the time required to calculate their forecasts is considerably lower than in the autoregressive models. Note that in order to avoid confusion between the $\alpha$ range on Rènyi entropy and the $\alpha$ parameter on Holt-Winters, henceforth $H_\alpha(X)$ is summarized as $H(X)$, and the coming alpha symbols will refer only to the forecasting adjustment.

The Holt-Winters model allows performing the prediction of the next observation $H(X)_{t+1}$ by analyzing three different components $B$, $T$ and $S$. These are defined in the following recursive expression:

$$B_t = \alpha(H_i - S_{t-N}) + (1 - \alpha)(B_{t-1} + T_{t-1}) \tag{5}$$

$$T_t = \beta(B_t - B_{t-1}) + (1 - \beta)T_{t-1} \tag{6}$$

$$S_t = \gamma(H_t - B_t) + (1 - \gamma)B_{t-n} \tag{7}$$

where $B_t$ is the base estimation at $t$, the estimation of the trend is $T_t$ and the estimation of the seasonal factor is $S_t$. The forecasting parameters $\alpha$, $\beta$ and $\gamma$ fall in the range $0 < \alpha, \beta, \gamma < 1$, and facilitate the adjustment of the smoothing. The prediction $H_{t+1}$ is usually calculated by additive or multiplicative operations. This approach considers the additive version, since it is assumed that the seasonal pattern of the series is independent of its trend. Consequently the forecast is calculated as follows:

$$H(X)_{t+1} = B_t + T_t + S_t \tag{8}$$

Another important aspect to keep in mind is the initialization method of $B_0$, $T_0$ and $S_0$ estimators. It is assumed that when no trend or seasonality is expected on the time series, the initialization of estimators based on the latest observations is preferable over the use of global measures. The implemented method is described in (Makridakis et al. (1998)), which has proven to behave particularly well in similar use cases. Namely, the last twenty-four observations are considered. The calculations performed are the following:

$$B_0 = \bar{M}_1 \tag{9}$$

$$T_0 = \frac{\bar{M}_2 - \bar{M}_1}{12} \tag{10}$$

$$S_{t-12} = \frac{p_t}{\bar{M}_1} \tag{11}$$

where $M_1$ summarizes the first twelve observations and $M_2$ the last dozen. The adjustment of parameters $\alpha$, $\beta$, $\gamma$ is obtained by calculating the values minimizing the function Sum of the Squared Errors of prediction (SSE), defined as:

$$SSE(\alpha, \beta, \gamma) = \sum_{t=1}^{N} (H(X)_t - H(X)_{t|t-1})^2 \tag{12}$$

### 4.3. Definition of prediction intervals

For the evaluation of the variation of entropy according with $X$ in the observation $t$, two thresholds are constructed: an upper threshold $Th_t$ and a lower threshold $Tl_t$. From these the prediction interval of the sensor is defined in the same way as is usually performed when implementing Holt-Winters (Hyndman et al. (2005)). They are expressed as follows:

$$Th_t(t) = p_0 + K \times \sqrt{var(E_t)} \tag{13}$$

$$Tl_t(t) = p_0 - K \times \sqrt{var(E_t)} \tag{14}$$

where $E_t$ is the prediction error in $t$ and $p_0$ is the prediction of the last observation. The prediction error is given by the difference between the forecast and the $t$ observation. The variance $Var(E_t)$ is calculated considering the prediction error of the last $t$ observations. In addition, the thresholds include a parameter $K$, from which it is possible to adjust the sensitivity of the detector. In the case of $D_H$ agents, the default value $Z = \frac{\alpha}{2}$ is assigned to $K$, thus relating the thresholds with the normal distribution of the series. Note that this is not a wrong decision considering publications as (Makridakis et al. (1998)), where it has shown that when the time series does not approach the normal distribution, the error is unrepresentative. Moreover the margin rate of both intervals is in the order $100(1 - \alpha)$.

On the other hand, as part of the adaptive response, the $D_A$ agents acquire the ability to increase their level of restriction based on the anomalous activities detected by $D_H$. Hence the parameter $k$ is determined by the following expression:

$$K(t) = K_{prev}(1 - \frac{Vol_{atk}}{Vol_{leg}}) \tag{15}$$

Where $K_{prev}$ is the last setting of $K$, $V_{atk}$ is the total volume of traffic monitored during the attack and $V_{leg}$ is the total volume of traffic monitored at the previous legitimate situation.

As an example, in Fig. 3, the time series associated with the entropy of the monitored traffic in part of one of the experiment and its prediction are shown. Legitimate traffic passes through the sensor until the observation at $t = 56$; then the injection of a large volume of traffic is observed. In Fig. 4, the prediction intervals of the sensor under the same circumstances are shown. When the attack is launched, both thresholds are exceeded. Then the agent reports of the incidence and drops the malicious packets, so the entropy back to their original values.

### 4.4. Identification of sources

Before the deployment of mitigation measures, and in order to allow specificity, it is important to identify the possible origins of the detected threat.

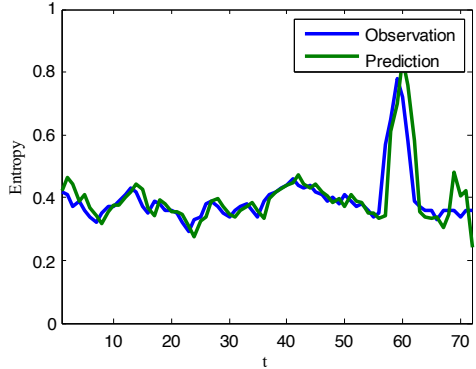

Figure 3: Example of forecasting of the entropy

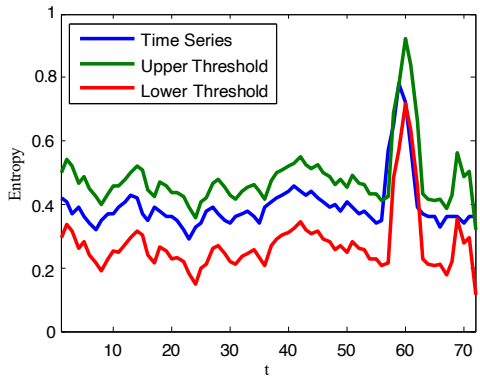

Figure 4: Example of variations on the thresholds of prediction

With this purpose, the different $p$ probabilities observed at $t$ are studied. The following assumptions are taken into account:

- Attacks could be originated from different IP addresses.

- Attacks may be directed against different destination IP addresses and ports.

- The routes Source, Destination, Port with higher $p$ are more likely to be involved in the attack

On this basis, the various instances of $X$ are grouped in function of their $p$. Given the nature of the information to be processed, the implemented algorithm must be unsupervised. Of the many options available, $k$-means algorithm has been chosen and the $K$ value has been adjusted by the *elbow* method. $K$-means was selected because is well known and it is especially resilient regarding the presence of outliers and errors in distance measurements. The *elbow* method compensates its main drawback: the need to previously specify the number of clusters to be generated. This method launches $k$-means with different $K$ values, until unrepresentative variations in the sum of the squared error between each member of the clusters with its central value occurs. Alternatives to this decision can be found in (Rui and Wunsch (2005)). At the end of the identification of sources, nodes grouped in the cluster with greater $p$ area tagged as suspicious, and therefore are quaran- tined.

## 5. Experimentation

In the implementation of the sensors, the observations are delimited by a fixed number of packets whose order of arrival is consecutive. A common alternative to this is their delimitation by considering time intervals. Both pose advantages and disadvantages, with the first choice being more efficient in the analysis of collections of previously captured traces, as is the case of most of the datasets analyzed (Ozcelik and Brooks (2015)). In addition, a sliding window of size $N$ that gathers the observations involved in the calculation of the entropy was applied. This boundedness is important to ensure that the implemented algorithms are computable, avoiding the case where $N \to \infty$. The proposal has been evaluated in two stages. Firstly, the accuracy of the various agents when dealing with DDoS flooding attacks was measured. On the other hand, several features related to the effectiveness of the deployment of the AIS were studied. The following describes each of them in detail.

### 5.1. Assessment of accuracy of the immune agents

Given the controversy relating to the evaluation methods of the effectiveness of intrusion detection system for identification of DDoS attacks, the scheme proposed in (Kumar and Selvakumar (2013)) was applied. This involves the use of two well-known datasets: KDD'99 (DoS (2016e)) and CAIDA'07 (DoS (2016b)); and in the generation of flooding attacks with the tool DDoSIM (DoS (2016c)) in a real use case. The following describes the principal characteristics of each test and their application.

### 5.1.1. KDD'99

The KDD'99 (DoS (2016e)) is one of the most referenced methodologies in the bibliography, and according to (Bhatia et al. (2014)), possibly the only one that presents a dataset with reliable labeling. It was created in 1999, under the KDD Cup competition, and from captures of traffic provided by DARPA'98. The competition task was to build a network intrusion detec- tor, a predictive model capable of distinguishing between "bad" connections, called intrusions or attacks, and "good" normal connections. It provides 41 different features of legitimate and malicious traffic samples. The attacks fall into four main categories: DoS (Denial of Service, e.g. *syn flood* ), R2L (unauthorized access from a remote machine, e.g. *guessing password* ), U2R (unauthorized access to local superuser (root) privileges, e.g., *buffer overflow* attacks) and probing (surveillance and other probing, e.g., *port scanning*). Originally it separates a subset of the datasets for the training stages of ex- pert intrusion detection systems, and the rest for their evaluation. Since the proposed system requires no training, all samples have been applied in the evaluation process, with the exception of the first observations, necessary for initialization of the predictive models. It is noteworthy that the antiq- uity of the dataset and the discovery of irregularities in its content have led to its discrediting. An important part of the research community considers KDD'99 unrepresentative, mainly due to lack of heterogeneity in comparison with current networks, old class of intrusions, errors when data gathering, etc. As is discussed in (Viswanathan et al. (2013)), this leads to the mistake of consider that the experimental results are scalable to real monitoring en- vironments. But despite this, it remains one of the most used methodologies, mainly due to the administrative difficulties associated with the publication of new datasets and the fact that it was implemented in most of the previous proposals.

### 5.1.2. CAIDA'07/08

The CAIDA'07 dataset (DoS (2016b)) provided samples of traffic traces containing DDoS flooding attacks (mainly ICMP, SYN and HTTP) moni- tored in August 2007. They are divided into files with extension pcap and spaced at time intervals of five minutes. As described in their documenta- tion, after the capture, most of the non-malicious contents were removed. Therefore it provides a good battery of tests to evaluate the effectiveness of the detection systems when analyzing flooding attacks, thus allowing their hit rates to be calculated. However, it is also necessary to determine their behavior when processing legitimate traffic, thereby calculating the false positive rates. Consequently, the use of CAIDA'07 is often complemented with the passive collection of traffic traces CAIDA'08 (DoS (2016a)), as described in (Kumar and Selvakumar 2013)). The latter compiles usual and legiti- mate traffic monitored at the data centers *Equinix* of San Jose and Chicago (the same networks as CAIDA'07). Both represent the traditional evaluation scheme with greater similitude to recent networks, allowing the contrast of the results, and involving a more current context. In the evaluation of the proposal, were separated into traces of 15,000 packets, and the distinction of the class of flooding attacks has not been taken into account.

### 5.1.3. DDoSIM and UCM traffic

This test scenario is a real use case. It combines the study of the habitual traffic on the subnet corresponding with the Faculty of Computer Science of the Complutense University of Madrid (UCM), with the analysis of flooding attacks injected by the tool DDoSIM (DoS (2016c)). The generated attacks act at application layer, and are based on the massive send of different HTTP and TCP requests. During its course, DDoSIM simulates the behavior of various zombie computers using random assignment of IP addresses, which are able to log into the victim servers. Once established, it proceeds to the flooding of requests. The captured traffic has been divided into two groups: legitimate and malicious. Both contain samples with traces of 40,000 packets in format pcap. The first one is applied to calculate the false positive rate of the approach. The other is to determine the hit rate.

### 5.2. Evaluation of the Artificial Immune System

To evaluate the effectiveness of the deployment of the AIS, a simulator capable of generating traffic distributions and different networks with different locations of $D_H$ and $D_A$ has been implemented. This is because none of the functional standards for the evaluation of similar systems provides a complete knowledge of the organization of various networks. In the generation of new networks, several parameters had been taken into account: number of nodes, legitimate traffic volume, branching component, and cyclic component. The last two determine the number of connections associated with each node and the number of cycles in the network when it is plotted as a graph of finite dimensions. On the other hand a tool for simulating flooding attacks has been developed. Given a network built by the previously described scheme, the tool emulates the injection of a certain amount of packets from a source node to a victim side, according with (Bhatia et al. (2014)).

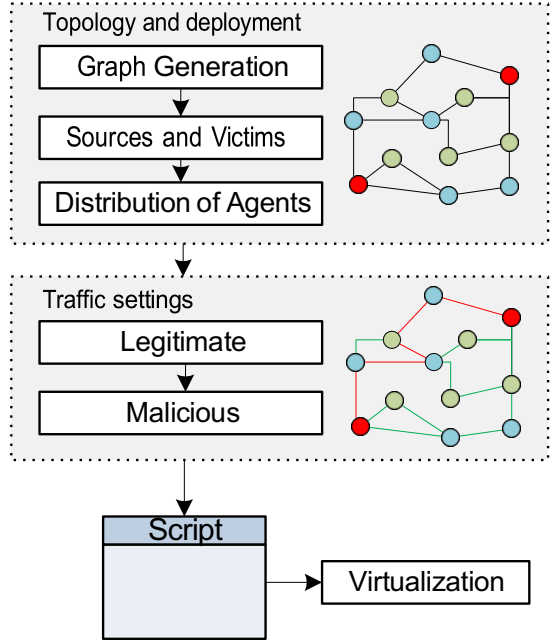

Figure 5: Construction of virtual networks for testing

In Fig. 5 a summary of the tasks involved in the creation of each of the networks in the experiments is shown. In a first step, the network topology and the distribution of agents are defined. The network is built according to the previously described parameters, and is represented by a graph where the vertices are its nodes, and the edges are its connections. Then the origin and the destination of the attacks are defined. The location of immune agents is defined by greedy graph coloring (Galinier and Hertz (2006)), where the two most frequent calculated colors represent the class of actors. Thus the amount of $D_A$ sensors dependent of each $D_H$ is regulated. The second level defined how the traffic is generated. The legitimate communications are randomly decided by taking into account the simulation parameters. The injection of traffic is carried out by the tool hping3. In the case of clean traffic, various protocols (FTP, HTTP, ICMP, etc.) and actions (transfer of files, requests, session maintenance, etc.) are performed. With all of this, a script that allows the deployment of the network in a virtualized environment is built. Through its execution it is possible to assess the effectiveness of the AIS in the test. In the experiments, 220 different networks have been considered. For each of them, the behavior of denial of service attacks with different power, directed between each possible pair of source and destination nodes had been studied. Different features of the proposal have been evaluated in function of the location of the immune agents, the flooding power of the attacks, the congestion of the network, or the malicious traffic mitigated.

## 6. Results

The results obtained when assessing the behavior of the agents separately and in the study of their capacity of collaboration according with the proposed AIS, are described below.

### 6.1. Detection of threats

At the evaluation of the effectiveness of the artificial immune agents when detecting threats, the adjustment parameters of the detectors are the variable $\alpha$ that defines the order of the traffic entropy, and the number of packets per observation. Variations on the first of them have a similar result with the three benchmarks: When higher is the value, the higher the level of restriction of the sensor. In such a way that when $\alpha > 3$, the deployment of the AIS is counterproductive because the false positive rate becomes excessively high (exceeding 20%). In general terms, the smaller the value of $\alpha$, the lower the rate of false positives. For this reason it was considered $\alpha = 1$ along the experimentation. Bearing this in mind, the following describes and discusses the results obtained by analyzing KDD'99, CAIDA'07/08 and DDoSIM injection against habitual UCM traffic.

### 6.1.1. KDD'99

In Fig. 6 the results obtained by analyzing the collection KDD'99 are shown. The X axis displays the number of packets per observation, whereas the Y axis indicates the True Positive Rate (TPR) and False Positive Rate (FPR). The hit rate has remained nearly unchanged in all the tests. However, the amount of errors in analyzing legitimate traffic is especially sensitive to the position in X; with fewer packets per sample, the system behaves more restrictively. Its ability in detection ceases to depend on it from 2,000 observations, at which it operates in saturation mode. This reason has led to study in more detail the two class of flooding attacks contained in more than 100,000 packages within the dataset: *smurf* and *neptune*. The rest (*back*, *teardrop*, *pod* and *land* ) are not taken into account because KDD'99 does not provide enough information to initialize the predictive models in the cases where there are lots of packages per observation. The true positive rates in saturation are 98.66% (*neptune*) and 100% (*smurf* ), with a false positive rate of 1.42%.

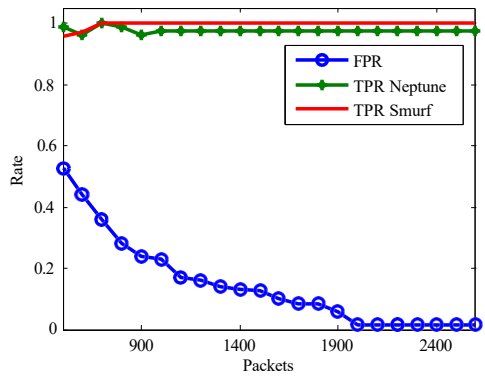

Figure 6: Results when analyzing KDD'99

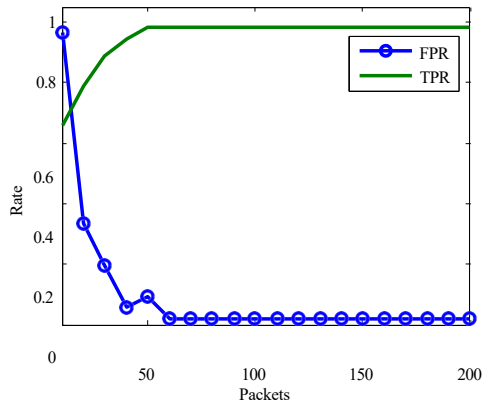

Figure 7: Results when analyzing CAIDA'07 and CAIDA'08

### 6.1.2. *CAIDA'07/08*

In Fig. 7 the results obtained when analyzing the traces of CAIDA'07 and CAIDA'08 are shown. The behavior of the agents is very similar to the previous test. But this time, the detectors operate in saturation with fewer packets per observation; in particular, around 160. In this case the hit rate is 98.11% and false positive is almost 1.91%. The difference between the results obtained in KDD'99 and CAIDA'07/08 is mainly due to the variations in the homogeneity of samples. In KDD'99 the traffic is older, and the scenario where the samples were gathered is more limited. Consequently, the legitimate samples are more like each other, thus reducing the tendency of the system to issue false positives.

### 6.1.3. DDoSIM and UCM traffic

In Fig. 8 the results obtained in the experimentation with UCM traffic and DDoSIM are shown. They remain the behavior of the previous tests, reaching saturation in 2,000 packets per observation. However, the accuracy obtained is considerably worse: the hit rate is 92.3% and the false positive rate is 8.3%. The difference precision achieved demonstrates that the good results obtained when applying the evaluation functional standards are not scalable to real networks. This is because the homogeneity of the traffic analyzed is much higher, according to the current usage models. But despite the high false positive rate, the quality of service of the protected environment will not be affected. The proportional increase in the rigor with which act the agents of the adaptive immune response and the specificity of the approach, allow that most of the legitimate traffic involved in the emission of false positives reach their destination by alternative ways; this will be unusual when dealing with malicious traffic.

The experimentation also emphasizes another important feature of the proposed method: most of the identified attacks have triggered alerts with proximity to the beginning and end of the malicious flooding. These are the observations where variations of entropy differ most from the legitimate traffic. When the attack is constant (as example, due to high rate attacks), the entropy tends to stabilize again, becoming invisible to the detector. Such behavior is illustrated more clearly in Fig. 9. It shows the impact on the traffic entropy of a distributed attack generated by DDoSIM against the UCM network. The X axis displays the observations and the Y axis indicates the value of the normalized entropy. The attack starts from the observation 60. Entropy anomalies are particularly visible on the following 15-20 observations, where the forecast exceeds the prediction thresholds, and thus the sensor reports an incidence. If mitigation measures (such as the packet drop applied in Fig. 2) are not adopted, the entropy is stabilized, albeit with much higher values. Most likely the attack will not be visible again until completion. At that time 15-20 observations reveal the descent of the entropy to their usual values. In view of this behavior, and in order to prevent evasion strategies, the combination of detection and mitigation measures is required, as is driven by the proposed AIS.

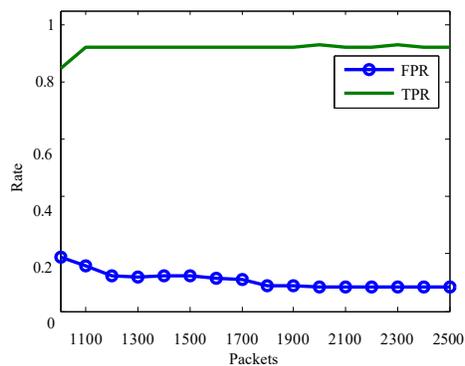

Figure 8: Results when analyzing traffic from DDOSIM and UCM

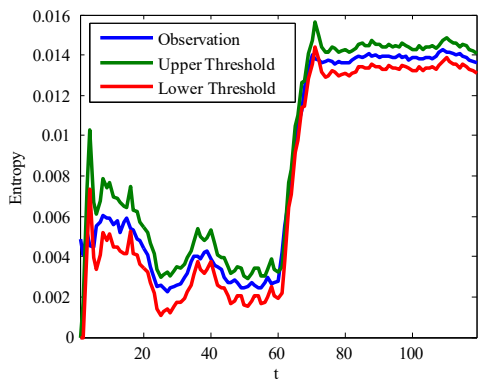

Figure 9: Evolution of the entropy with DDOSIM and UCM

## 6.2. Artificial immune reactions

The behavior of the proposed systems taking into account the location of the agents, the flooding power of the attack, the legitimate congestion of the protected network and the mitigation capacity of the AIS are described below.

### 6.2.1. Location of the immune agents

In Fig. 10 the results when $D_H$ and $D_A$ are triggered in different placements are shown. Y axis indicates the TPR/FPR and the X axis the location of the agents. The latter is indicated by a value in range 0 to 1, which determines the distance in the path between source and victim, where 0 is the location of the source, and 1 the location of the victim.

When new attacks are launched, the average TPR is 0.85 and the FPR is 0.072. There is a trend: near the ends, the TPR value approaches 1. However, at intermediate points the accuracy is reduced by up TPR is 0.70. This occurs in the points closest to the equidistant location between the ends. When the $D_A$ agents are activated by the immune adaptive response, the trend is repeated. Nevertheless the average TPR was increased by 7%. The greatest improvement is observed at intermediate distances. The worst TPR value is 0.82, which implies the improvement of 12%. The adaptive immunity response has low impact on the FPR, i.e., it increased 0.6% at the worst case.

In view of these results, and assuming that most of the conventional IPS behave similarly to the innate response, it can be stated that the pro- posed AIS poses a significant improvement in the cases where they find more difficulties to operate adequately. This occurs at the nodes above halfway between the attacker and the victim, mainly because the malicious traffic can spread over a larger number of alternative paths; near the ends they tend to converge (close to the victim) or diverge (close to the attacker).

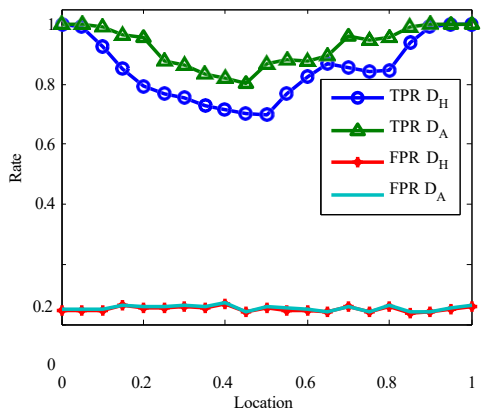

Figure 10: Results based on the location of the immune agents

### 6.2.2. Flooding Power

In Fig. 11 the accuracy of the system depending on the flooding potency is observed. The Y axis details the TPR/FPR and the X axis indicates the power of the attacks. The last is represented in values in the range of 0 to 1, which correspond to the percentage of the bandwidth that aims to occupy, with 0 being no traffic injection and 1 means the complete saturation of the connections. As shown, when the attack is more powerful than 0.5, the accuracy in both responses is similar and close to 1. Therefore, changes are irrelevant. However, when less noisy attacks, they are more difficult to be detected. These cases are where the adaptive response improves the performance of the system. At the best case, the TPR has increased by 26.5% in the power range 0.3 to 0.4. In summary, the higher the power of the attacks, the greater the noise caused, and therefore the threats are easier to detect. When attacks are less visible, a more significant improvement of the proposed AIS on the conventional IPS is observed.

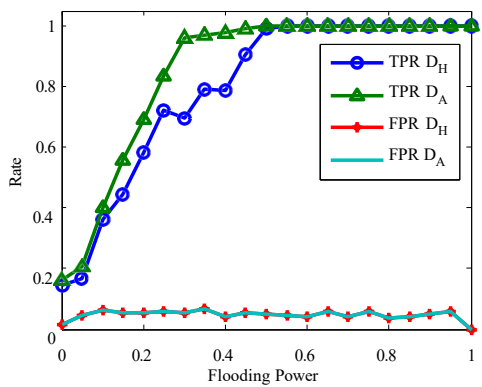

Figure 11: Results based on the flooding attack

### 6.2.3. Network Congestion

In Fig. 12 the detection capability in function of the volume of legiti- mate traffic flowing through the networks is shown. The X axis details the TPR/FPF, and the Y axis indicated the legitimate traffic volume. The last is indicated with values in range 0 to 1. As it was done in the above test, X represents the saturation level of the network. Results are very similar to those in Fig. 11. When the traffic density is low, the proposed strategy is very accurate, as is the case of the conventional schemes. This is because the attacks are much more visible, taking a larger share of bandwidth. However, when congestion is high, especially above 0.7, the hit rate decreases, and the false positive rate increases. In the same way as in the previous tests, this problem is reduced by the activation of the adaptive response. Particularly, this strategy results in an improvement of the TPR of 13.7% when greatest slope (congestion between 0.6 and 0.7).

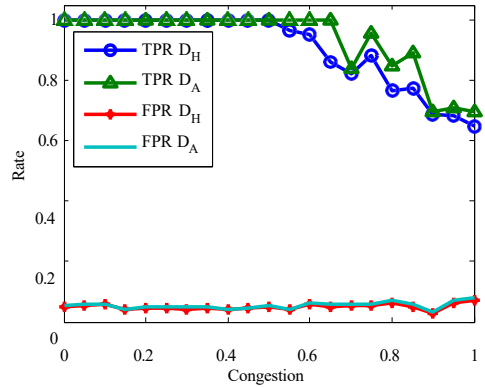

Figure 12: Results based on the legitimate network congestion

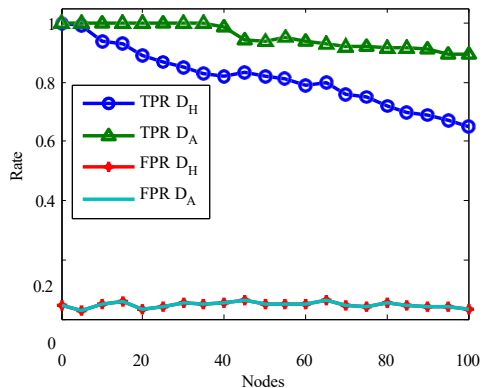

Figure 13: Mitigation of attacks based on the number of compromised nodes

### 6.2.4. Mitigation

In Fig. 13 the rate of attacks that have not reached their destinations, depending on the number of nodes in their paths, is observed. When acting solely the innate response, 81.4% of the threats had been blocked at the worst case. However, the adaptive response was able to prevent 95.5% of them, i.e., has improved accuracy by 14.1%. From the figure it also follows that the longer the path, the greater the probability of success. This is because the load of the attack can be distributed more effectively. Bearing this in mind, it is possible to state that the larger the protected network, more relevant is the improvement obtained with this proposal. This is because the attacks that traverse larger networks often flow from greater amounts of nodes. A great effect is not achieved by avoiding their pass through few connections if there are many other routes available, as occur in the conventional IPS. But the adaptive response has the ability to spread rapidly over the network, thus increasing security measures in almost all available paths.

## 7. Conclusions

In this paper a novel approach for detecting and mitigating DoS flooding attacks based on the emulation of the behavior of the immune system of the human beings has been proposed. It implied the design of different artificial immune agents and their distribution throw the protected network.

The system has been evaluated in two stages. Firstly, the accuracy of the detection methods was measured taking into account the functional standards KDD'99, CAIDA'07 and recent traffic of the UCM network compromised by the tool DDoSIM. The results obtained were satisfactory, empowering their collaboratively deployment. On the other hand, their efficiency as AIS has been evaluated. This has entailed its implementation on differ- ent virtualized networks and the assessment of their effectiveness based on different parameters. Regardless of the criteria from which the behavior of the AIS has been evaluated (location of the immune agents, flooding power, network congestion or mitigation), the adaptive response has always shown more effectiveness than the innate response. This results in a significant improvement in their ability to detect and mitigate attacks, without penalization in their error rates when processing legitimate traffic. The innate response behaves in the same way as the conventional IPS, so in the adap- tive reactions it is possible to study the raw effectiveness of the approach over the conventional mitigation schemes. Bearing these in mind, it is possible to confirm that the emulation of the biological immune responses is a very good way to enhance the classical countermeasures against DoS attacks.

In view of these results, this proposal is promoting the initialization of new lines of research. The simplest of them is based on the improvement of metrics and forecasting methods. Others are introducing the addition of novel immune agents, thus allowing the AIS to better perform in more complex use cases. But undoubtedly the most interesting are those that focus on its deployment at real uses cases. Any work in this regard will be especially interesting, given that the evaluation of the AIS has been mostly performed in simulated test scenarios. In particular, our research group is progressing towards its implementation on Software Defined Networks, and in adaptation to the anonymous network Tor.